\documentclass[12pt]{article}

\usepackage{amssymb, amsmath}
\include{amslatex}

\def\QED{\hskip0.1em\hfill\null\ \null\nobreak\hfill
\kern3pt\lower1.8pt\vbox{\hrule\hbox   {\vrule\kern1pt\vbox{\kern1.7pt
\hbox{$\scriptstyle   QED$}\kern0.2pt}\kern1pt\vrule}\hrule}}

\parskip=\medskipamount
\newcommand{\calC}{\mathcal{C}}

\newcommand{\calH}{\mathcal{H}}
\newcommand{\calI}{\mathcal{I}}

\newcommand{\calP}{\mathcal{P}}
\newcommand{\calS}{\mathcal{S}}

\newcommand{\calQ}{\mathcal{Q}}
\newcommand{\calU}{\mathcal{U}}

\newcommand{\bfR}{\mathbb{R}}

\newcommand{\be}{\begin{equation}}
\newcommand{\ee}{\end{equation}}
\newcommand{\bea}{\begin{eqnarray*}}
\newcommand{\eea}{\end{eqnarray*}}

\newtheorem{definition}{Definition}[section]
\newtheorem{lemma}[definition]{Lemma}
\newtheorem{theorem}[definition]{Theorem}
\newtheorem{prop}[definition]{Proposition}

\newtheorem{remark}[definition]{Remark}

\newenvironment{mproof}{\textbf{Proof:}\,}

\newcommand{\arxiv}[1]{\textit{#1}}

\newcommand{\norm}[1]{\left\Vert{#1}\right\Vert}

\newcommand{\vf}{\mathfrak{X}}
\newcommand{\proj}{\mathbf{h}}

\newcommand{\im}{\mathrm{Im}}

\newcommand{\tD}{\tilde{D}}
\newcommand{\tF}{\tilde{F}}
\newcommand{\tX}{\tilde{X}}
\newcommand{\tY}{\tilde{Y}}
\newcommand{\tZ}{\tilde{Z}}

\newcommand{\talpha}{\tilde{\alpha}}
\newcommand{\teta}{\tilde{\eta}}
\newcommand{\tOmega}{\tilde{\Omega}}
\newcommand{\tpi}{\tilde{\pi}}
\newcommand{\tS}{\tilde{S}}
\newcommand{\tcalC}{\tilde{\calC}}

\newcommand{\tcalP}{\tilde{\calP}}
\newcommand{\tcalQ}{\tilde{\calQ}}

\title{Geometric aspects of nonholonomic field theories}
\author{J. Vankerschaver\thanks{Research Assistant of the Research
    Foundation--Flanders (FWO--Vlaanderen)}, F. Cantrijn \\
    {\small Department of Mathematical Physics and Astronomy, Ghent
      University} \\
    {\small Krijgslaan 281, B--9000 Ghent (Belgium)} \\
    {\small e-mail: Joris.Vankerschaver@UGent.be, Frans.Cantrijn@UGent.be}
    \\ and \\
    M. de Le\'on, D. Mart\'{\i}n de Diego \\
    {\small Instituto de Matem\'aticas y F\'{\i}sica Fundamental, CSIC} \\
    {\small C/ Serrano 123, E--28006 Madrid (Spain)} \\
    {\small e-mail: mdeleon@imaff.cfmac.csic.es, d.martin@imaff.cfmac.csic.es}}

\begin{document}
\maketitle
\begin{abstract}
A geometric model for nonholonomic Lagrangian field theory is
studied. The multisymplectic approach to such a theory as well as
the corresponding Cauchy formalism are discussed. It is shown that
in both formulations, the relevant equations for the constrained
system can be recovered by a suitable projection of the equations
for the underlying free (i.e.\ unconstrained) Lagrangian system.
\end{abstract}
\section{Introduction}
During the past decades, much effort has been devoted to the
differential geometric treatment of mechanical systems subjected
to nonholonomic (i.e.\ velocity-dependent) constraints. To a large
extent the growing interest in this field has been stimulated by
its close connection to problems in control theory (see for
instance the recent books \cite{bloch,cortes}, which also contain
extended bibliographies on the subject). As far as the study of
Lagrangian (or Hamiltonian) systems with nonholonomic constraints
is concerned, one can essentially distinguish between two
different approaches. The first one, commonly called
``nonholonomic mechanics", is based on d'Alembert's principle and
on an additional rule that specifies a bundle of covectors along
the constraint submanifold, representing the admissible reaction
forces (or, alternatively, a subbundle of the tangent bundle,
representing the admissible infinitesimal virtual displacements).
The second one is a constrained variational approach, often coined
``vakonomic mechanics". As is well-known, the dynamical equations
generated by both approaches are in general not equivalent (for
that matter, see for instance \cite{arnold,cortes2,lewis}).

In this paper we will study an extension of nonholonomic mechanics
to the treatment of (classical) Lagrangian field theories with
external constraints. For the time being, we will confine
ourselves to first-order field theories with constraints that
depend on the independent variables (say the space-time
coordinates), the fields and their first-order partial derivatives
only. For the sake of clarity it should be emphasized that we use
the term ``classical field theory" here in a broad sense, i.e. it
essentially refers to any physical (or other) system that can be
described by the Euler-Lagrange equations derived from a
Lagrangian density. The constraints involved are called `external'
to distinguish them from the type of constraints appearing in
gauge theories that find their origin in the possible degeneracy
of the Lagrangian.

Constrained field theories have already been studied extensively
in the literature but, as far as the case of external constraints
is concerned, those treatments are usually based on a
field-theoretic analogue of vakonomic mechanics (see e.g.\
\cite{MPSW}). The nonholonomic approach we are going to discuss
here is a continuation of some work by E. Binz \textit{et al.}
\cite{nhfields02}. We will refer to this theory as ``nonholonomic
field theory" mainly for its formal analogy with nonholonomic
mechanics and also to distinguish it from the constrained
variational approach. However, the term nonholonomic should
perhaps be used here with some caution. Indeed, as can be inferred
from a remark in \cite{MPSW}, there does not seem to exist a
general agreement yet on the precise meaning of the notion of
nonholonomic constraint in field theory. In that respect, a recent
paper by O. Krupkova  \cite{krupkova} may come to the rescue with
a view on straightening out this matter, at least from a purely
mathematical point of view.

The mathematical framework for a nonholonomic field theory that
has been proposed in \cite{nhfields02} involves, among others, a
generalization of d'Alembert's principle and of the so-called
Chetaev rule that is commonly used in nonholonomic mechanics to
characterize the bundle of constraint forms representing the
admissible reaction forces. The constrained field equations, as
well as an extension of the so-called De Donder-Weyl equations for
classical field theories, are then derived in a finite-dimensional
multisymplectic setting. The treatment in \cite{nhfields02} also
briefly deals with the Cauchy formalism for nonholonomic field
theory.

The purpose of the present paper is to continue and extend the
work described in \cite{nhfields02}. First of all, we will relax
the Chetaev-type conditions by allowing for constraint forms that
need not be determined by the constraints themselves. Next, in the
multisymplectic setting we will construct a kind of projection
operator that maps solutions of the free (i.e.\ unconstrained) De
Donder-Weyl equations into solutions of the modified (constrained)
De Donder-Weyl  equations that have been proposed in
\cite{nhfields02}. This is similar to the situation encountered in
nonholonomic mechanics (see e.g. \cite{autnonhol99}). Finally, we
also treat the Cauchy formalism for nonholonomic field theory. In
particular, we will show that in case a global `space-time'
splitting of the base manifold can be fixed, the resulting
structures and equations on the infinite-dimensional space of
Cauchy data are reminiscent of those appearing in the geometric
treatment of time-dependent nonholonomic mechanics. This is in
full agreement with the results described for the unconstrained
case by A. Santamar\'{\i}a \cite{santamaria04} (see also
\cite{symm04}).

The scheme of the paper is as follows. In the next section we
recall some basic elements from the multisymplectic approach to
(unconstrained) Lagrangian field theories on jet bundles, mainly
in order to fix some of the notation that will be used. In
Sections 3 and 4 we then discuss the construction of a
nonholonomic model for first-order Lagrangian field theories with
external constraints. The corresponding constrained field
equations are given, as well as the modified De Donder-Weyl
equations. Next, a projector is constructed that maps solutions of
the De Donder-Weyl equations for the unconstrained Lagrangian
system into solutions of the constrained De Donder-Weyl equations.
In Section 5, as an example to illustrate the theory, we briefly
consider the case of incompressible hydrodynamics. Section 6 is
devoted to the Cauchy formalism for nonholonomic field theory and
in Section 7 we conclude with some general comments.

\section{Multisymplectic approach to Lagrangian field
theories}\label{multisymplectic}

There is an extended literature on the multisymplectic approach to
classical field theories: see for instance
\cite{firstorder91,gimmsyI,overview02}, where the interested
reader can also find further references. Before reviewing some
aspects of that theory, we first briefly recall some basic notions
from jet bundle theory for which we mainly rely on the book by
D.J. Saunders \cite{saunders89}.
\subsection{Jet bundles}\label{jets}
Consider a fibre bundle $\pi: Y \rightarrow X$ of rank $m$, whose
base space $X$ is assumed to be an oriented manifold of dimension
$n + 1$, equipped with a fixed volume form $\eta$. (All manifolds,
maps, vector fields and differential forms are assumed to be of
class $C^{\infty}$). The first-order jet bundle $J^{1}\pi$ is the
set of equivalence classes $j^{1}_x\phi$ consisting of those
(local) sections $\phi$ of $\pi$ around a point $x \in X$ having
the same Taylor expansion up to order one at $x$. $J^{1}\pi$ is a
$(n+1+m+(n+1)m)$-dimensional manifold, fibred over $X$ with
projection $\pi_1: J^{1}\pi \longrightarrow X$. In addition,
$J^1\pi$ has the structure of an affine bundle over $Y$ of rank
$(n+1)m$, modelled on the vector bundle $V\pi\otimes \pi^*(T^*X)
\longrightarrow Y$, with $V\pi$ the bundle of $\pi$-vertical
tangent vectors to $Y$, and with projection denoted by $\pi_{1,0}:
J^1\pi \longrightarrow Y$. In particular, we have $\pi_1 = \pi
\circ \pi_{1,0}$.

As far as coordinates on $Y$ are concerned, we will always consider
local bundle coordinates adapted to $\pi$, written as $(x^\mu, y^a)$,
$\mu=1,\ldots, n+1; a = 1, \ldots, m$, and where the coordinate system
$(x^\mu)$ on $X$ is taken such that $\eta = d^{n+1} x: = dx^1 \wedge
\cdots \wedge dx^{n+1}$. The induced bundle coordinates on $J^1\pi$
will then be denoted by $(x^\mu, y^a, y^a_\mu)$.

Given a vector field $\xi$ on $Y$, its first jet prolongation to
$J^1\pi$ will be written as $\xi^{(1)}$. In coordinates, if
\[ \xi = \xi^{\mu}(x,y)\frac{\partial}{\partial x^{\mu}} +
\xi^a(x,y)\frac{\partial}{\partial y^a}\, ,
\]
then
\begin{equation}\label{prolong}
\xi^{(1)} = \xi^{\mu}\frac{\partial}{\partial x^{\mu}} +
\xi^a\frac{\partial}{\partial y^a} + \left(\frac{d\xi^a}{dx^{\mu}}
-
y^a_{\nu}\frac{d\xi^{\nu}}{dx^{\mu}}\right)\frac{\partial}{\partial
y^a_{\mu}}\,.
\end{equation}

In terms of the volume form $\eta$ on $X$, one can construct a
special tensor field $S_\eta$ on $J^1\pi$, called the
\emph{vertical endomorphism} (see \cite{saunders89} for a formal
definition). It is a vector-valued $(n+1)$-form that has the
following expression in the coordinate system described above:
$$
  S_\eta = (dy^a - y^a_\nu dx^\nu) \wedge d^{n}x_\mu
    \otimes \frac{\partial}{\partial y^a_\mu}\,,
$$
where $\displaystyle{d^{n}x_{\mu}=i_{\frac{\partial}{\partial
x_{\mu}}}d^{n+1}x}$. Note in passing that we will not make a
notational distinction between the volume form $\eta$ on $X$ and
its pull-back to $Y$ or to $J^1\pi$ under the respective
projections.

A 1-form $\theta\in \Lambda^1(J^1\pi)$ is said to be a contact
1-form whenever $(j^1\phi)^*\theta=0$ for each section $\phi$ of
$\pi$. Locally, the module of contact forms is spanned by the
1-forms $ \theta^a= dy^a-y^a_{\mu}\, dx^{\mu} $.

In this paper, we will frequently deal with connections (in the
sense of Ehresmann) on $\pi_1$.  Such a connection induces in
particular a direct sum decomposition $TJ^1\pi = H\pi_1 \oplus
V\pi_1$, where $V\pi_1$ is the bundle of $\pi_1$-vertical tangent
vectors to $J^1\pi$ and $H\pi_1$ is  a complementary distribution,
called the horizontal distribution. The corresponding horizontal
projection operator will always be denoted by $\proj$. We recall
that a connection on $J^1\pi$ uniquely determines a section
$\Upsilon$ of the bundle $(\pi_1)_{1, 0}: J^1\pi_1 \rightarrow
J^1\pi$, called a \emph{jet field}, and that, conversely, each jet
field characterizes a unique connection on $\pi_1$ (see
\cite{saunders89} for details). Therefore, in the sequel we will
use the denominations `connection' (on a jet bundle) and `jet
field' interchangeably.

An \emph{integral section} of a jet field $\Upsilon: J^1\pi
\rightarrow J^1\pi_1$ is a local section $\sigma$ of $\pi_1$ such
that $j^1 \sigma = \Upsilon \circ \sigma$. A jet field, or the
associated connection, is called \emph{semi-holonomic} if its
integral sections $\sigma$ (if they exist!) are first jet
prolongations of sections of $\pi$, i.e. $\sigma = j^1\phi$ for
some $\phi: X \longrightarrow Y$. Whenever a semi-holonomic jet
field admits integral sections, it is called \emph{holonomic} (or
integrable). We now recall the conditions, in coordinates, for a
jet field to be (semi-)holonomic.

In coordinates, the horizontal projector $\proj$ of a connection
on $\pi_1$ can be written as
\begin{equation}\label{connection}
  \proj = dx^\mu \otimes \left( \frac{\partial}{\partial x^\mu} +
  \Gamma^a_\mu \frac{\partial}{\partial y^a} + \Gamma^a_{\mu\nu}
  \frac{\partial}{\partial y^a_\nu}\right),
\end{equation}
for some functions $\Gamma^a_\mu (x^\kappa,y^b,y^b_\kappa)$ and
$\Gamma^a_{\mu\nu}(x^\kappa,y^b,y^b_\kappa)$, and the associated
jet field $\Upsilon$ then reads:
\[
  \Upsilon: J^1\pi \longrightarrow J^1\pi_1,\;(x^\mu, y^a, y^a_\mu)
  \mapsto(x^\mu, y^a, y^a_\mu,
  \Gamma^a_\mu, \Gamma^a_{\mu\nu}).
\]
A local section $\sigma$ of $\pi$, with $\sigma(x) = (x^\mu,
\sigma^a(x), \sigma^a_\mu(x))$, is an integral section of the jet
field $\Upsilon$ if
$$
  \frac{\partial \sigma^a}{\partial x^\mu} = \Gamma^a_\mu \quad
  \text{and} \quad \frac{\partial \sigma^a_\mu}{\partial x^\nu} =
  \Gamma^a_{\mu\nu}.
$$
From this expression, it easily follows that the connection will
be semi-holonomic if $\Gamma^a_\mu = y^a_\mu$ and holonomic (or
integrable) if, in addition, $\Gamma^a_{\mu\nu} =
\Gamma^a_{\nu\mu}$.

We wish to emphasize that the previous discussion about
connections on jet bundles can be presented in fully intrinsic
terms: we refer again to \cite{saunders89} for details. In
particular, the condition for a connection (or jet field), with
horizontal projector $\proj$, to be semi-holonomic, can be
expressed by demanding that
\begin{equation} \label{charsemihol}
  i_{\proj} \theta = 0, \quad \hbox{ for each contact $1$-form } \theta.
\end{equation}
We will make use of this characterization later on.

\subsection{First-order Lagrangian field
theories}\label{lagrfieldtheory} Given a fibre bundle $\pi: Y
\longrightarrow X$ and its associated first-order jet bundle
$J^1\pi$, as considered above, we now briefly recall some aspects
from the multisymplectic formulation of first-order Lagrangian
field theory on $J^1\pi$, where the fields are the (local)
sections of $\pi$. Consider a Lagrangian density $L\eta$ with $L$
a smooth function (the Lagrangian) defined on $J^1\pi$. We say
that $L$ is \emph{regular} if its Hessian matrix is
non-degenerate, i.e.
$$
  \det\left(\frac{\partial^2 L}{\partial y^a_\mu \partial
  y^b_\nu}\right)\neq 0
$$
at each point of $J^1\pi$.

Using the vertical endomorphism $S_\eta$ we can construct the
following $(n+1)$-form on $J^1\pi$:
\[ \Theta_L = S_\eta^\ast dL + L \eta \]
and we then define the $(n + 2)$-form $\Omega_L = -d\Theta_L$,
called the \emph{Poincar\'e-Cartan form}.  The following
coordinate expression for $\Omega_L$ will often be convenient:
\begin{equation}\label{omega}
  \Omega_L = - \frac{\partial L}{\partial y^a}dy^a \wedge d^{n+1}x
    -d\left(\frac{\partial L}{\partial y^a_\mu}\right) \wedge
    (dy^a - y^a_\nu dx^\nu) \wedge d^{n}x_\mu.
\end{equation}
If $L$ is regular, which we will always assume in the sequel, the
Poincar\'e-Cartan form is a multisymplectic form according to the
following definition.
\begin{definition}[see \cite{msmanifolds99,tulczyjew03,symm04}] \label{def:ms}
  A closed $m$-form $\Omega$ on a manifold $M$ is called
  \emph{multisymplectic} if the mapping $v_x \in T_x M \mapsto i_{v_x}
  \Omega(x) \in \Lambda^{m-1}(T^*_xM)$ is injective for all $x \in M$.
\end{definition}
Note that symplectic forms ($m=2$) and volume forms ($m=\dim M$)
are particular examples of multisymplectic forms and, moreover,
these are the only two cases where the mappings in
definition~\ref{def:ms} are surjective as well as injective
(assuming $M$ is finite dimensional).

We now give a brief outline of the derivation of the
Euler-Lagrange equations for first-order field theories. For more
detailed treatments we refer to \cite{bsf88, overview02,
saunders89}. The action functional associated to the given
Lagrangian density $L\eta$ is defined as
\begin{equation}\label{action}
  \calS(\phi) = \int_U (j^1 \phi)^\ast L\eta
\end{equation}
where $U$ is an open subset of $X$ with compact closure, and $\phi$ is
a section of $\pi$ defined over $U$.  A section $\phi$ is an
\emph{extremal} of (\ref{action}) if
$$
  \frac{d}{dt} \calS(\varphi_t \circ \phi) \Big|_{t = 0} = 0,
$$
for any local flow  $\{\varphi_t\}$ on $Y$, consisting of
diffeomorphisms defined on a neighborhood of $\phi (U)$,
satisfying $\pi \circ \varphi_t = \pi$, $\varphi_0 = \text{id.}$,
and keeping the boundary of $\phi(U)$ fixed. Extremal sections are
characterized by the following property.
\begin{prop}
  A (local) section $\phi$ of $\pi$ is an extremal for the action (\ref{action})
  if and only if
  \begin{equation}\label{globalEL}
    (j^1\phi)^\ast (i_{\zeta} \Omega_L) = 0,
  \end{equation}
  for all vector fields $\zeta$ on $J^1\pi$.
\end{prop}
\begin{mproof}
See for instance \cite[prop. 7.1.2]{bsf88}, \cite[thm.
3.1]{gimmsyI} or \cite{firstorder91}.\QED
\end{mproof}

In coordinates, equation (\ref{globalEL}) yields the well-known
Euler-Lagrange equations:
$$
 \frac{\partial L}{\partial y^a} - \frac{d}{dx^\mu}
 \left( \frac{\partial L}{\partial y^a_\mu}\right)
  = 0 \quad (a=1,\ldots,m).
$$

A more general problem consists in looking for sections $\tau$ of
$\pi_1$ such that $\tau^*(i_{\zeta}\Omega_L) = 0$ for all vector
fields $\zeta$ on $J^1\pi$ (i.e.\ $\tau$ need not be the
prolongation of a section of $\pi$). This leads to the so-called
De Donder equations of Lagrangian field theory. In case of a
regular Lagrangian, the De Donder equations are equivalent to the
Euler-Lagrange equations (\ref{globalEL}). In this paper, we will
mostly be concerned with a kind of linearized version of these
equations that is more easy to handle and is obtained by looking
for a connection on $\pi_1$ whose integral sections will be
extremals of (\ref{action}). More precisely we have the following
important proposition:

\begin{prop}
  Given a holonomic connection with horizontal projector
  $\proj$, then the integral sections of the associated jet field
  are extremals of (\ref{action}) if and only if
  \begin{equation}\label{DDW}
    i_\proj \Omega_L = n \Omega_L\,.
  \end{equation}
\end{prop}
\begin{mproof}
  See \cite[thm. 5.5.5]{saunders89} and \cite{constraint94}.\QED
\end{mproof}

Moreover, a simple coordinate computation shows that, given a
regular Lagrangian $L$ on $J^1\pi$, a connection on $\pi_1$
satisfying (\ref{DDW}) will automatically be semi-holonomic.
Equation (\ref{DDW}) is also referred to as the \emph{De
Donder-Weyl equation} of Lagrangian field theory.

\section{Nonholonomic Lagrangian field theory} \label{par:nh}

We now bring constraints into the picture.  Suppose we have a
Lagrangian system on $J^1\pi$, with regular Lagrangian $L$. Let
$\calC \hookrightarrow J^1 \pi$ be a submanifold of $J^1\pi$ of
codimension $k$, representing some external constraints imposed on
the system. Although one can certainly consider more general
situations (see e.g. \cite{nhfields02}), for the sake of clarity,
we will confine ourselves to the case that $\calC$ projects onto
the whole of $Y$, i.e. $\pi_{1,0}(\calC) = Y$, and that the
restriction $(\pi_{1, 0})_{\vert\calC}: \calC \longrightarrow Y$
of $\pi_{1,0}$ to ${\cal C}$ is a (not necessarily affine) fibre
bundle. In particular the latter is a quite restrictive condition
but, with proper caution, one can probably carry out the further
analysis under some weaker assumption. Finally, one could still
require that ${\cal C}$ should not be the first jet bundle of a
subbundle of $\pi$, but since it does not really affect the
present treatment we will not insist on that (i.e.\ our discussion
also covers the case of ``holonomic" constraints that do not
essentially depend on the derivatives of the fields).

Since $\calC$ is a submanifold of $J^1\pi$, one may always find a
covering of $\calC$ consisting of open subsets $U$ of $J^1\pi$,
with $U \cap \calC \ne \varnothing$, such that on each $U \in
\calU$ there exist $k$ functionally independent smooth functions
$\varphi_\alpha$ that locally determine $\cal C$, i.e.
$$
  \calC \cap U = \{ \gamma \in J^1\pi : \varphi_\alpha(\gamma) = 0
    \text{ for $1 \le \alpha \le k$} \}.
$$
We remark here that the assumption that $(\pi_{1,0})_{|\calC}$ is
a fibre bundle implies, in particular, that the matrix
$\displaystyle{(\frac{\partial
  \varphi_\alpha}{\partial y^a_\mu})(\gamma)}$ has maximal rank $k$
at each point $\gamma \in {\cal C} \cap U$.

With the given data --- a Lagrangian and a constraint submanifold
$\cal C$ --- the question now arises how to construct a suitable
constrained field theory. As pointed out in the introduction, one
possibility is to follow a constrained variational approach as,
for instance, in \cite{MPSW}. Inspired by the situation in
classical mechanics one may also think of another approach, called
``nonholonomic", which involves some additional ingredients (see
\cite{nhfields02}). For mechanical systems with nonholonomic
constraints it is now well-known that the ``vakonomic" and the
``nonholonomic" equations of motion are not equivalent in general.
Equivalence is achieved, however, if the constraint functions can
be written as total time derivatives of velocity-independent
constraints, i.e.\ if we are basically dealing with holonomic
constraints (see e.g. \cite{cortes2} and references therein).

\subsection{The constraint distribution}\label{constrdistr}

Making a small digression to nonholonomic mechanics, we recall
that the construction of the equations of motion of a mechanical
system with nonholonomic constraints is based on the so-called
d'Alembert principle and involves, among others, the specification
of a suitable bundle of admissible ``reaction forces'' (and a
corresponding bundle of admissible virtual velocities), defined
along the constraint submanifold. This choice relies on an
additional rule or principle. In nonholonomic mechanics it is
quite common to use the so-called Chetaev principle, whereby the
bundle of reaction forces is constructed directly in terms of the
given constraints. In principle, however, the specification of the
appropriate bundle of reaction forces (or virtual displacements),
compatible with the given constraints, is problem dependent and
need not necessarily be based on Chetaev's rule. For a critical
discussion of this matter we refer to \cite{marle98}; see also
\cite{terra}.

Returning to the case of first-order field theory with external
constraints, we now introduce a special subbundle $F$ of rank $k$
of the bundle of exterior $(n+1)$-forms on $J^1\pi$ defined along
the constraint submanifold $\cal C$, where we recall that $k$ is
the codimension of ${\cal C}$. This bundle, which we will simply
refer to as the bundle of constraint forms, will play a role
similar to that of the bundle of reaction forces in nonholonomic
mechanics. We define this bundle by considering a submodule ${\cal
F}$ of rank $k$ of the module of $(n+1)$-forms $\Phi$, defined on
a neighborhood of ${\cal C}$, and which are $n$-horizontal and
$1$-contact, i.e.\ $\Phi$ vanishes when contracted with any two
$\pi_1$-vertical vector fields and $(j^1\phi)^*\Phi = 0$ for any
section $\phi$ of $\pi$. In particular, one can find an open cover
${\cal U}$ of ${\cal C}$ such that on each open set $U \in {\cal
U}$, the module ${\cal F}$ is generated by $k$ independent
$(n+1)$-forms $\Phi_{\alpha}$ that locally read
\begin{equation}\label{constraintform}
  \Phi_\alpha = (C_\alpha)^\mu_a (dy^a - y^a_\nu dx^\nu) \wedge d^n
  x_\mu = (C_\alpha)^\mu_a \theta^a \wedge d^n x_\mu\,,
\end{equation}
for some smooth functions $(C_{\alpha})^\mu_a$ on $U$.
Independence of the forms $\Phi_{\alpha}$ clearly implies that the
$(k \times (n+1)m)$-matrix whose elements are the
$(C_{\alpha})_a^\mu$, has constant maximal rank $k$. The
\emph{bundle of constraint forms} is then defined by
\[
F= \bigcup_{\gamma \in\, {\cal C}}F_{\gamma}\, \quad \mbox{with\ }
F_{\gamma}=\{\Phi(\gamma)\,|\, \Phi \in {\cal F}\}\,.
\]
At this point, the reason for selecting a constraint bundle of the
type described above is primarily based on the analogy with
nonholonomic mechanics.
\begin{remark}\label{chetaev} In \cite{nhfields02} the authors have
constructed the bundle of constraint forms by considering a
natural extension of the Chetaev-principle that is commonly used
in mechanics when dealing with nonlinear nonholonomic constraints.
More precisely, they define the local generators $\Phi_{\alpha}$
of the bundle of constraint forms by putting
\[
\Phi_{\alpha}:= S_{\eta}^*(d\varphi_{\alpha})\,,
\]
where the $\varphi_{\alpha}$ are the local constraint functions
(see the beginning of Section 3). One easily verifies that these
$\Phi_{\alpha}$ are indeed of the form (\ref{constraintform}),
with $(C_{\alpha})_a^{\mu} = \frac{\partial
\varphi_{\alpha}}{\partial y^{a}_{\mu}}$. In the case we are
considering, the independence of these $\Phi_{\alpha}$ is
guaranteed by our initial assumption that ${\cal C}$ should have a
fibre bundle structure over $Y$.
\end{remark}

As we will now show, the constraint bundle $F$ gives rise to a
distribution $D$ along $\cal C$, called the \emph{constraint
distribution}. As above, consider an open cover ${\cal U}$ of
$\cal C$ such that on each $U \in {\cal U}$, the module $\cal F$
is generated by $k$ independent $(n+1)$-forms $\Phi_{\alpha}$ of
the form (\ref{constraintform}).

\begin{prop}\label{prop:constrdistr}
For each $\alpha$, there exists a unique vector field
$\zeta_\alpha \in \vf(U)$ such that
  \begin{equation}\label{symplectic}
    i_{\zeta_\alpha} \Omega_L = -\Phi_\alpha.
  \end{equation}
\end{prop}
\begin{mproof}
Take $\zeta_\alpha$ to be a $\pi_{1,0}$-vertical vector field on
$U$, i.e. $\zeta_\alpha = (\zeta_\alpha)^a_\mu
\frac{\partial}{\partial y^a_\mu}$.  Herewith, equation
(\ref{symplectic}) reduces to
\begin{equation}\label{constrforce}
  (\zeta_\alpha)^a_\mu \frac{\partial^2 L}{\partial y^a_\mu \partial
  y^b_\nu} = (C_\alpha)^\nu_b,
\end{equation}
which determines the $(\zeta_\alpha)^a_\mu$ uniquely, as $L$ is
supposed to be regular.  This already proves the existence of a
solution of (\ref{symplectic}). Uniqueness then follows from the
fact that $\Omega_L$ is multisymplectic.\QED
\end{mproof}

The vector fields $\zeta_\alpha$ span a $k$-dimensional
distribution $D_U$ on $U$.  It is not difficult to check that for
any two open sets $U,V \in {\cal U}$ with nonempty intersection,
and for each $\gamma \in U \cap V$, $D_U(\gamma) = D_V(\gamma)$.
Indeed, assume $\cal F$ is generated on $U$ by $k$ independent
forms $\Phi_{\alpha}$ and on $V$ by $k$ independent forms
$\overline{\Phi}_{\alpha}$, then there exists a nonsingular matrix
of functions $r^{\beta}_{\alpha}$ on $U \cap V$ such that
$$
\Phi_{\alpha} = r^{\beta}_{\alpha}\overline{\Phi}_{\beta}\,.
$$
If we denote the corresponding generators of $D_U$ by
$\zeta_{\alpha}$ and those of $D_V$ by $\overline{\zeta}_\alpha$,
it readily follows from the previous proposition that
$$
  {{\zeta}_\alpha}_{|U\cap V} = r_\alpha^\beta {\overline{\zeta}_\beta}_{|U\cap V}\,,
$$
which proves that $D_U = D_V$ on $U \cap V$. Consequently, the
local distributions described in the previous proposition induce a
well-defined (global) distribution $D$ along the constraint
submanifold ${\cal C}$, whose sections are $\pi_{1,0}$-vertical
vector fields. Moreover, using a similar argument as above one
easily verifies that this distribution does not depend on the
initial choice we made for an open cover $\cal U$ of $\cal C$.

\subsection{The nonholonomic field equations}
Summarizing the above, we are looking for a field theory built on
the following data: (i) a Lagrangian density $L\eta$ with regular
Lagrangian $L \in C^{\infty}(J^1\pi)$; (ii) a constraint
submanifold ${\cal C} \subset J^1\pi$ that can be locally
represented by equations of the form
$\varphi_{\alpha}(x^{\mu},y^a,y^a_{\mu})=0$, for
$\alpha=1,\ldots,k$ and where the matrix $({\partial
  \varphi_\alpha}/{\partial y^a_\mu})$ has maximal rank $k$;
(iii) a bundle $F$ of constraint forms and an induced constraint
distribution $D$, both defined along ${\cal C}$, whereby $F$ is
locally generated by $k$ independent $(n+1)$-forms
(\ref{constraintform}), and $D$ is defined according to the
construction described in Proposition \ref{prop:constrdistr}.

To complete our model for nonholonomic field theory, we now have
to specify what the field equations are. Proceeding along the same
lines as in \cite{nhfields02} we introduce the following
definition, using a generalization of d'Alembert's principle.

\begin{definition} A (local) section $\sigma$ of $\pi: Y \longrightarrow
X$, defined on an open set $U \subset X$ with compact closure, is
a solution of the constrained problem under consideration if
$j^1\sigma (U) \subset {\cal C}$ and
\[
\int_U (j^1\sigma)^*{\cal L}_{\xi^{(1)}} L\eta = 0\,,
\]
for all $\pi$-vertical vector fields $\xi$ on $Y$ that vanish on
the boundary of $\sigma(U)$ and such that
\[i_{\xi^{(1)}}\Phi = 0\quad (*)
\]
for all sections $\Phi$ of the bundle $F$ of constraint forms.
\end{definition}
Putting $\xi = \xi^a(x,y)\partial/\partial y^a$ and taking into
account the expression (\ref{prolong}) for the prolonged vector
field $\xi^{(1)}$, it is easily seen that the condition $(*)$
translates into
\[ (C_{\alpha})^{\mu}_a\xi^a = 0\,,
\]
where the $(C_{\alpha})^{\mu}_a$ are the coefficients of the
constraint forms introduced in (\ref{constraintform}). One can
then verify that if $\sigma(x) = (x^{\mu}, \sigma^a(x))$ is a
solution of the constrained problem, then the functions
$\sigma^a(x)$ satisfy the following system of partial differential
equations
\begin{eqnarray}\label{nonholfieldeqns}
\frac{\partial L}{\partial y^a} - \frac{d}{dx^\mu}
\left(\frac{\partial L}{\partial y^a_\mu}\right)&=&
   \lambda^{\alpha}_{\mu}(C_\alpha)^{\mu}_a \quad
  (a=1, \ldots,m)\,,\\
\varphi_{\alpha}(x^\mu,\sigma^a(x),\frac{\partial
\sigma^a}{\partial x^\mu}(x))&=&0 \quad (\alpha = 1, \ldots, k)\,.
\end{eqnarray}
As usual, the (a priori) unknown functions
$\lambda^{\alpha}_{\mu}$ play the role of `Lagrangian
multipliers'. The equations (\ref{nonholfieldeqns}) are called the
\emph{nonholonomic field equations} for the constrained problem.
Note that if the bundle $F$ of constraint forms is defined
according to a Chetaev-type prescription (see Remark 3.1), then we
recover the nonholonomic field equations derived in
\cite{nhfields02}.

Let ${\cal I}(F)$ be the ideal of differential forms, defined
along ${\cal C}$, generated by the constraint forms: i.e\ any
element of ${\cal I}(F)$ is of the form $\sum_i \lambda_i \wedge \Phi^i$, for
some $\Phi^i \in {\cal F}$ and arbitrary differential forms
$\lambda_i$. Again proceeding along the same lines as in
\cite{nhfields02} we can formulate the following modification of
the De Donder-Weyl problem for nonholonomic Lagrangian field
theory: find a connection on $\pi_1: J^1\pi \longrightarrow X$
with horizontal projector $\proj$ such that along the constraint
submanifold $\cal C$
\begin{equation}\label{nonholddw}
i_{\proj}\Omega_L - n\Omega_L \in {\cal I}(F) \quad \mbox{and}
\quad \mbox{Im\ } \proj \subset T{\cal C}\,.
\end{equation}
For simplicity we will refer to (\ref{nonholddw}) as the
\emph{nonholonomic De Donder-Weyl equation}. In coordinates, if we
represent $\proj$ by (\ref{connection}) one can easily check that
the relation on the left of (\ref{nonholddw}) leads to the
following set of equations for the connection coefficients of the
connection we are looking for:
\begin{eqnarray*}
(\Gamma^b_{\nu} - y^b_{\nu})\left(\frac{\partial^2 L}{\partial
y^a_{\mu}\partial y^b_{\nu}}\right)&=&0\,,\\
\frac{\partial L}{\partial y^a} - \frac{\partial^2 L}{\partial
x^{\tau}\partial y^a_{\tau}} - \Gamma_{\tau}^b\frac{\partial^2
L}{\partial y^b \partial y^a_{\tau}} - \Gamma^b_{\tau
\nu}\frac{\partial^2 L}{\partial y^b_{\tau} \partial y^a_{\nu}} +
(\Gamma^b_{\nu} - y^b_{\nu})\frac{\partial^2 L}{\partial y^a
\partial y^b_{\nu}}&=& \lambda^{\alpha}_{\tau}(C_\alpha)^{\tau}_a\,,
\end{eqnarray*}
for $a=1, \ldots, m$ and $\mu = 1, \ldots, n+1$ and some
Lagrangian multipliers $\lambda^{\alpha}_{\tau}$. This should
still be supplemented by the requirement that for any $\gamma \in
{\cal C}$ and any $v \in T_{\gamma}J^1\pi$, $\proj (v) \in
T_{\gamma}{\cal C}$. This is equivalent to requiring that
${\proj}(v)(\varphi_{\alpha}) =0$ for all $v \in T_{\cal
C}J^1\pi$, where $\varphi_{\alpha}\, (\alpha = 1, \ldots ,k)$ are
the (local) constraint functions. If, locally, $\proj$ is written
in the form (\ref{connection}), then the previous condition
translates into the following additional equations for the
connection coefficients in points of $\cal C$:
\[
\frac{\partial \varphi_\alpha}{\partial x^\mu} +
\Gamma_{\mu}^b\frac{\partial \varphi_\alpha}{\partial y^b} +
\Gamma^b_{\mu \nu}\frac{\partial \varphi_\alpha}{\partial y^b_\nu}
= 0\quad \mbox{for all} \quad \mu = 1, \ldots, n+1;\; \alpha= 1,
\ldots ,k.
\]
One can prove that in case of a regular Lagrangian, integral
sections of a connection satisfying (\ref{nonholddw}) will be
1-jet prolongations of solutions of the nonholonomic field
equations (see \cite{nhfields02} for details).

\section{The nonholonomic projector} \label{par:nhproj}

The purpose of the present section is to show that for a
nonholonomic first-order field theory in the sense described
above, one can construct, under an appropriate additional
condition, a projection operator which maps solutions of the De
Donder-Weyl equation (\ref{DDW}) for the free (i.e. unconstrained)
Lagrangian problem into solutions of the nonholonomic De
Donder-Weyl equation (\ref{nonholddw}).

Given a constrained problem as described in the previous section,
with regular Lagrangian $L$, constraint manifold ${\cal C} \subset
J^1\pi$ and constraint distribution $D$, we now impose the
following \emph{compatibility condition}: for each $\gamma \in
\calC$
\begin{equation}\label{compatibility}
    D(\gamma) \cap T_\gamma \calC = \{ 0 \}.
\end{equation}
If ${\cal C}$ is locally defined by $k$ equations
$\varphi_{\alpha}(x^{\mu},y^a,y^a_{\mu})=0$ and if $D$ is locally
generated by the vector fields $\zeta_{\alpha}$ (see subsection
\ref{constrdistr}), a straightforward computation shows that the
compatibility condition is satisfied iff
\[
\det\left(\zeta_{\alpha}(\varphi_{\beta})(\gamma)\right) \neq 0\,,
\]
at each point $\gamma \in {\cal C}$. Indeed, take $v \in T_\gamma
\calC \cap D(\gamma)$.  Then $v = v^\alpha \zeta_\alpha(\gamma)$,
for some coefficients $v^\alpha$.  On the other hand, $0 =
v(\varphi_\beta) = v^\alpha \zeta_\alpha(\varphi_\beta)(\gamma)$.
Hence, if the matrix
$\left(\zeta_\alpha(\varphi_\beta)(\gamma)\right)$ is invertible,
we may conclude that $v = 0$ and the compatibility condition
holds. The proof of the converse is similar.

We now have the following result.
\begin{prop}\label{dirsum}
If  the compatibility condition (\ref{compatibility}) holds, then
at each point $\gamma \in \calC$ we have the decomposition
  $$
    T_\gamma J^1\pi = T_\gamma \calC \oplus D(\gamma).
  $$
\end{prop}
\begin{mproof}
The proof immediately follows from (\ref{compatibility}) and a
simple counting of dimensions: $\dim T_\gamma \calC = \dim
T_\gamma J^1\pi - k$ and $\dim D(\gamma) = k$.\QED
\end{mproof}

The direct sum decomposition of $T_\calC J^1\pi$ determines two
complementary projection operators $\calP$ and $\calQ $:
$$
  \calP: T_\calC J^1\pi \rightarrow T\calC \quad \text{ and } \quad
  \calQ = I - \calP: T_\calC J^1\pi \rightarrow D\,,
$$
where $I$ is the identity on $T_{\cal C}J^1\pi$. We will call
$\cal P$ \emph{the nonholonomic projector} associated to the given
constrained problem.

Given a connection on $\pi_1$ such that the associated horizontal
projector $\proj$ is a solution of the free De Donder-Weyl
equation (\ref{DDW}), we will prove that the operator $\calP \circ
{\proj}_{|T_{\cal C}J^1\pi}$ satisfies the constrained De
Donder-Weyl equation (\ref{nonholddw}). Note that this operator is
only defined along $\calC$ and, therefore, strictly speaking it is
not the horizontal projector of a connection on $\pi_1$. However,
one can show that its restriction to $T{\cal C}$ induces a genuine
connection on the restricted bundle $(\pi_1)_\calC: {\cal C}
\longrightarrow X$, and so the constrained De Donder-Weyl equation
still makes sense for this kind of map.

\begin{lemma} \label{lemma:connection}
The map ${\cal P} \circ {\proj}_{|T_{\cal C}J^1\pi}: T_{\cal
C}J^1\pi \longrightarrow T{\cal C}\,(\subset T_{\cal C}J^1\pi), v
\longmapsto {\cal P}({\proj}(v))$ is a projector whose restriction
${\proj}_{\cal P}$ to $T{\cal C}$ induces a connection on
$(\pi_1)_{|\calC}: {\cal C} \longrightarrow X$.
\end{lemma}
\begin{mproof}
First of all, we check that for each $\gamma \in \calC$ the map
$\calP_\gamma \circ \proj_\gamma$ is a projector.  Indeed, taking
into account that $\im \calQ = D$ is $\pi_{1,0}$-vertical, it
follows that for all $v \in T_\gamma J^1\pi$
$$ (\proj_\gamma \circ \calP_\gamma)(v) = \proj_\gamma (v) -
(\proj_\gamma\circ \calQ_\gamma)(v) = \proj_\gamma (v).
$$
and therefore
$$
  \left( \calP_\gamma \circ \proj_\gamma \right)^2 =
      \calP_\gamma \circ \proj_\gamma.
$$
The restriction ${\proj}_{\cal P}$ of ${\cal P} \circ
{\proj}_{|T_{\cal C}J^1\pi}$ to $T{\cal C}$ obviously is still a
projector. The key point we now have to prove is that $\im
({\proj}_{\cal P})$ is a complementary bundle to
$V(\pi_1)_{|\calC}$ in $T{\cal C}$, i.e.
\begin{equation}\label{constrconn}
  \im ({\proj}_{\cal P}) \oplus V(\pi_1)_{|\calC} = T\calC\,.
\end{equation}
For that purpose we start by observing that along $\calC$ we have
$T\calC \cap V\pi_1 = V(\pi_1)_{|\calC}$. In view of Proposition
\ref{dirsum} one can then easily derive the following direct sum
decomposition:
\begin{equation}\label{verticals}
  V(\pi_1)_{|\calC} \oplus D = V\pi_1 \quad \text{(along $\calC$)}.
\end{equation}
Next, by taking into account the fact that the constraint
distribution $D$ is vertical, and therefore that
$\proj_\calP(T_\gamma\calC) = (\calP \circ \proj)(T_\gamma
J^1\pi)$ for every $\gamma \in \calC$, it is a routine exercise to
verify that
\begin{equation}\label{horspaces}
\dim (\calP \circ \proj)(T_\gamma J^1 \pi) = \dim \proj(T_\gamma
J^1\pi)\,.
\end{equation}

We now prove the direct sum decomposition (\ref{constrconn}). Take
any $ v \in T{\cal C}$ with $v \in \im(\proj_{\cal P}) \cap
V(\pi_1)_{|\cal C}$, then there exists a vector $w \in T{\cal C}$
such that $v = \calP(\proj(w)) = \proj(w) - \calQ(\proj(w))$.
Since $v$ is $\pi_1$-vertical, we conclude that $\proj(w) = 0$
and, hence, $v = 0$. This already implies that $\im(\proj_{\cal
P}) \cap V(\pi_1)_{|\calC} = 0$. The equality (\ref{constrconn})
now follows from a simple dimensional argument. Indeed, relying on
Proposition \ref{dirsum} as well as on (\ref{verticals}) and
(\ref{horspaces}), we have at each point $\gamma \in {\cal C}$:
\begin{eqnarray*}
  \dim (\proj_{\cal P}(T_{\gamma}{\cal C})) + \dim
  V_{\gamma}(\pi_1)_{|\calC}
    & = & \dim ({\proj}(T_{\gamma}J^1\pi)) + \dim {V_{\gamma}\pi_1} - \dim D(\gamma) \\
    & = & \dim (T_{\gamma}J^1\pi) - \dim D(\gamma) \\
    & = & \dim T_{\gamma}\calC\,.
\end{eqnarray*}
This concludes the proof that ${\proj}_{\cal P}= \calP \circ
\proj_{|T{\cal C}}$ is the horizontal projector of a connection on
$(\pi_1)_{|\calC}$.\QED
\end{mproof}

Although $(\pi_1)_{|\cal C}: {\cal C} \longrightarrow X$ is not a
first-order jet bundle, we will say that a connection on
$(\pi_1)_{|\calC}$, with associated horizontal projector
$\hat{\proj}$, is \emph{semi-holonomic} if for each contact
$1$-form $\theta$ on $J^1\pi$
\begin{equation}\label{semihol}
  i_{\hat{\proj}} j^\ast \theta = 0,
\end{equation}
where $j: \calC \hookrightarrow J^1\pi$ is the canonical
injection. Suppose $\tau: X \longrightarrow {\cal C}$ is an
integral section of a connection on $(\pi_1)_{|{\cal C}}$, in the
sense that $T\tau(T_xX) \subset \hat{\proj}(T_{\tau(x)}{\cal C})$
for all $x \in \mbox{Dom\ } \tau$. Then, if the given connection
is semi-holonomic one can verify that, locally, $\tau$ can be
written as the first jet prolongation of a (local) section of
$\pi$.

As mentioned at the end of subsection~\ref{lagrfieldtheory}, the
regularity of $L$ together with the fact that $\proj$ satisfies
the free De Donder-Weyl equation, imply that $\proj$ is a
semi-holonomic connection on $J^1\pi$. Herewith one can prove the
following result.

\begin{lemma} \label{lemma:semihol}
The connection on $(\pi_1)_{|\calC}$ defined in Lemma
  \ref{lemma:connection}, with horizontal projector ${\proj}_{\cal P}$,
is semi-holonomic.
\end{lemma}
\begin{mproof}
We will use the fact that $\proj$ is semi-holonomic and therefore
satisfies (\ref{charsemihol}). Let $v \in T_{\gamma}J^1\pi$ be a
$\pi_{1, 0}$-vertical vector, then for any contact $1$-form
$\theta$ on $J^1\pi$ we have that $i_v \theta(\gamma)  = 0$. Now,
for each $v \in T_{\cal C}J^1\pi$ we have that $({\cal P}\circ
{\proj} - {\proj})(v) = - {\cal Q}({\proj}(v)) \in D$ and, hence,
$({\cal P}\circ {\proj} - {\proj})(v)$ is ${\pi}_{1,0}$-vertical.
Therefore $i_{\calP \circ
  \proj} \theta (v) = i_\proj \theta (v) = 0$ for any contact $1$-form
$\theta$ and any $v \in T_{\cal C}J^1\pi$.  From this one can
readily deduce that ${\proj}_{\cal P}$ satisfies (\ref{semihol})
and so we may conclude that the induced connection on
$(\pi_1)_{|\calC}$ is indeed semi-holonomic.\QED
\end{mproof}

We now arrive at the main result of this section. From now on, for
ease of notation, we will use the projector ${\cal P} \circ
{\proj}$ without further indication of its domain. The latter
should be clear from the context.

\begin{theorem} Consider a constrained problem of the type described
above, with regular Lagrangian $L$, constraint submanifold ${\cal
C} \subset J^1\pi$ and bundle of constraint forms $F$, and assume
the compatibility condition (\ref{compatibility}) holds. Let
$\proj$ be the horizontal projector of a connection on $\pi_1$,
satisfying the free De Donder-Weyl equation (\ref{DDW}) and let
$\calP$ be the nonholonomic projector associated to the
constrained problem. Then, the projector ${\cal P} \circ \proj$
determines a solution of the constrained De Donder-Weyl problem
(\ref{nonholddw}) and restricts to the horizontal projector of a
semi-holonomic connection on $(\pi_1)_{|\cal C}: {\cal C}
\longrightarrow X$.
\end{theorem}
\begin{mproof}
Along ${\cal C}$ we can rewrite the free De Donder-Weyl equation
as
$$
  i_{\calP \circ \proj} \Omega_L - n\Omega_L = -i_{\calQ \circ
    \proj} \Omega_L.
$$
Therefore, in order to prove that ${\cal P} \circ {\proj}$
satisfies the constrained De Donder-Weyl equation, we only need to
verify that the right-hand side is an element of $\calI(F)$.

We can write the projector ${\proj}$ as $\proj = dx^\mu \otimes
H_\mu$, with $H_\mu = (\partial/\partial x^\mu)+
\Gamma^a_\mu(\partial/\partial y^a) + \Gamma^a_{\mu
\nu}(\partial/\partial y^a_\nu)$ (cf.\ (\ref{connection})). Along
$\calC$ we can then put $\calQ(H_\mu) = \lambda^\alpha_\mu
\zeta_\alpha$ for some functions $\lambda^\alpha_\mu$ and with the
vector fields $\zeta_\alpha$ as defined in Proposition
\ref{prop:constrdistr}. Then, at each point $\gamma \in {\cal C}$ and
for any $v_1, \ldots, v_{n + 2} \in T_{\gamma}J^1\pi$ we obtain
\begin{eqnarray*} (i_{\calQ \circ \proj}
  \Omega_L)(v_1, \ldots, v_{n + 2}) & = & \sum_{i = 1}^{n + 2}
  (-1)^{i+1} \Omega_L((\calQ \circ \proj)(v_i), v_1, \ldots,
  \hat{v}_i, \ldots, v_{n + 2}) \\
    & = & \sum_{i = 1}^{n + 2}
  (-1)^{i+1} \lambda^\alpha_\mu dx^\mu(v_i) (i_{\zeta_\alpha} \Omega_L)
    (v_1, \ldots, \hat{v}_i, \ldots, v_{n + 2}) \\
    & = & - \lambda^\alpha_\mu(dx^\mu \wedge \Phi_\alpha)(v_1, \ldots,
  v_{n + 2})\,.
\end{eqnarray*}
This shows that, along ${\cal C}$,
$$
  i_{\calQ \circ \proj} \Omega_L = -\lambda^\alpha_\mu dx^\mu \wedge
  \Phi_\alpha \in \calI(F),
$$
which completes the proof of the first part of the theorem.

The proof that ${\cal P} \circ {\proj}$ induces a semi-holonomic
connection on $(\pi_1)_{|\cal C}$ follows from the previous lemmas
\ref{lemma:connection} and \ref{lemma:semihol}. \QED
\end{mproof}

Note that even in case a connection on $\pi_1$, with horizontal
projector ${\proj}$ satisfying the free De Donder-Weyl equation,
is holonomic (or integrable), the `projected' semi-holonomic
connection ${\proj}_{\cal P}= \calP \circ \proj$ on $(\pi_1)_{\cal
C}$ need not admit integral sections in general.

\section{An example: incompressible hydrodynamics}

As an example of a field theory with an external constraint, we
consider the case of an incompressible fluid flow.  This problem
has already been treated for instance in \cite{MPSW}, using the
constrained variational approach. From the point of view of
``nonholonomic field theory" this is perhaps an a-typical example
since, as we shall see, the constrained field equations resulting
from the nonholonomic approach are essentially the same as those
derived in \cite{MPSW}. The reason for this probably stems from
the fact that, as we will show, the incompressibility constraint
can be written as a divergence. Recall that for a mechanical
system with a nonholonomic constraint that arises from a total
time derivative of a function on the configuration space, the
nonholonomic and the vakonomic equations of motion are equivalent.

\subsection{The constrained problem}
We will consider a simplified model of incompressible fluid motion
in the sense that we will not bother about the technicalities
related to precise domain and boundary conditions (see \cite{MPSW}
for a more detailed description of the geometric model). We
identify the base space $X$ with $\bfR \times \bfR^3 \equiv
\bfR^4$, with coordinates $(x^\mu)=(t,x^i)$ representing time $t=
x^0$ and the material variables $x^i$ ($i=1,2,3$). The volume form
on $X$ is the standard Euclidean volume $\eta:= d^4x = dt\wedge
dx^1 \wedge dx^2 \wedge dx^3$. For the total space we take $Y = X
\times \bfR^3$ with coordinates $(t, x^i, y^a)$. The projection
$\pi: Y \longrightarrow X$ then reads $\pi (t,x^i,y^a) = (t,x^i)$.
In order to preserve some consistency with \cite{MPSW}, we will
denote the corresponding bundle coordinates on $J^1\pi$ by $(t,
x^i, y^a, v^a_0, v^a_i)$.
\begin{remark}  In continuum mechanics it is common to denote
the coordinates on $X$ and $Y$ by $(t, X^I)$ and $(t, X^I, x^i)$,
respectively, and the coordinates on $J^1\pi$ by $(t, X^I, x^i,
v^i, F^i_I)$.  However, we will not follow that convention here.
\end{remark}
In addition, we will equip the fibres $\bfR^3$ of $X$ (over the
time axis) and of $Y$ (over $X$) with the standard Euclidean
metric, although one could replace them by more general Riemannian
manifolds (see \cite{MH,MPSW}).

A section $\phi (t,x^i) = (t,x^i, \phi^a(t,x))$ of $\pi$ can be
seen as a map taking a material point $x$ of the fluid and mapping
it at each time $t$ onto its position $\phi^a(t,x),\ (a=1,2,3)$ in
space. Following \cite{MPSW} we write the Lagrangian density as
\begin{equation} \label{lagrangian}
  L(\gamma)d^4x = \frac{1}{2} \norm{v_0}^2 \rho\, d^4x - W(v^a_i)
  \rho\, d^4x,
\end{equation}
where the function $\rho=\rho(x)$ represents the material density,
and $W$ is the stored energy function.  Note that $W$ depends only
on the $v^a_i$, i.e. the ``spatial'' jet bundle coordinates. Next,
we introduce the function ${\mathcal J}: J^1\pi \rightarrow \bfR$
given by
$$
 {\mathcal J}(\gamma) = \det \left(v^a_i(\gamma)\right).
$$
(note that $(v^a_i)$ is a square matrix). For any section $\phi$
of $\pi$, ${\mathcal J}\circ j^1\phi$ measures the volume change
of a small fluid element under the `flow' represented by $\phi$.
In particular, the incompressibility requirement can be expressed
by the condition ${\mathcal J}(j^1\phi) = 1$, i.e.\ we have the
constraint
\begin{equation} \label{incompressibility}
  \varphi(\gamma): = {\mathcal J}(\gamma) - 1 = 0\,,
\end{equation}
defining the constraint submanifold ${\cal C}$.

For the bundle $F$ of constraint forms we take the line-bundle
along ${\cal C}$, generated by the $4$-form
\begin{eqnarray*}
\Phi:&=& S^*_{\eta}(d\varphi)\\
& =& \frac{\partial \varphi}{\partial v^a_\mu}(dy^a - v^a_\nu
dx^\nu)\wedge d^3x_\mu\\
&=& {\mathcal J} (v^{-1})^i_a (dy^a - v^a_\nu dx^\nu)\wedge
d^3x_i\,,
\end{eqnarray*}
(i.e.\ we adopt the generalized Chetaev principle: see Remark
\ref{chetaev}).

\subsection{The nonholonomic field equations}
Before proceeding towards the field equations, we make the
additional assumption that we are dealing with a \emph{barotropic
fluid} which, in particular, implies that $W$ depends on the
$v^a_i$ through $\mathcal J$, i.e. $W = W({\mathcal J})$.
 The nonholonomic field equations (\ref{nonholfieldeqns}) for a
 barotropic fluid with Lagrangian
(\ref{lagrangian}), subject to the incompressibility constraint
(\ref{incompressibility}) and with constraint form $\Phi$, become
\[
\rho\, \delta_{ab}\frac{d}{dt}v^b_0  -
    \frac{\partial}{\partial x^j} \left(\rho W'
  {\mathcal J}(v^{-1})^j_a \right) = \lambda_i{\mathcal J}(v^{-1})^i_a \quad
  (a=1,2,3)\,,
\]
which should be considered together with the constraint equation
$\varphi(t,x^i,y^a,v^a_0,v^a_i) = {\mathcal J} -1 = 0$. This
should be compared with equation (4.8) in \cite{MPSW}. In that
paper, the the field equations for an incompressible barotropic
fluid were derived by means of a constrained variational approach.
Since there is only one constraint equation, this approach gives
rise to only one Lagrangian multiplier $P$, which can be
interpreted as pressure. If we put $\lambda_i = \frac{\partial
P}{\partial x^i}$, it is seen that, for the present example, the
nonholonomic equations and the constrained variational equations
are essentially the same. When thinking of the comparison between
nonholonomic and vakonomic mechanics, the reason for this is to be
found in the fact that the incompressibility constraint is
determined by a divergence. More precisely, we have the following
property.
\begin{prop}
The constraint function $\varphi$ can be written (locally) as a
total divergence, i.e.\ there exist functions $\psi^\mu$ such that
$\varphi = \frac{d\psi^\mu}{d x^\mu}$.
\end{prop}
\begin{mproof}
One can easily verify that
$$
  \frac{d}{d x^\mu} \left( \frac{\partial \varphi}{\partial v^a_\mu}
     \right) - \frac{\partial \varphi}{\partial v^a} \equiv 0,
$$
i.e. $\varphi$ is a ``null-Lagrangian", which is equivalent to
$\varphi$ being a divergence (see e.g. \cite[thm. 4.7]{olver86}).
More directly, if we consider the functions
$$
  \psi^0 = 0 \quad \text{and} \quad \psi^i = \frac{1}{3}{\mathcal J} y^a (v^{-1})_a^i -
  x^i\,,
$$
with $v^{-1}$ the inverse of the matrix $(v^a_i)$, which are
well-defined on a neighborhood of ${\cal C}$, a rather tedious but
straightforward computation shows that $\varphi =
d\psi^\mu/dx^\mu$. \QED
\end{mproof}
A detailed study of the comparison between the constrained
variational approach and the nonholonomic approach to constrained
field theories will be the subject of forthcoming work.

\subsection{The nonholonomic projector}
To illustrate some further concepts defined in the preceding
sections, we now turn to the explicit form of the nonholonomic
projector $\calP$ for the example of incompressible fluid (not
necessarily barotropic). As there is only one constraint, the
bundle of constraint forces $D$ is spanned by a single vector
field $\zeta = \zeta^a_\mu {\partial}/{\partial v^a_\mu}$. The
coefficients of this vector field can be derived from
(\ref{constrforce}) where, in the present case, $C_a^\mu =
{\partial \varphi}/{\partial v^a_\mu}$:
$$
  \left( \begin{array}{cc}
    1 &  0\\
    0 & \frac{\partial^2 W}{\partial v^a_i \partial v^b_j} \\
  \end{array} \right)
  \left( \begin{array}{c}
    \zeta^b_0 \\ \zeta^b_j \\
  \end{array} \right) = \left( \begin{array}{c}
    0 \\ {\mathcal J}(\gamma)\, (v^{-1})^i_a \end{array} \right).
$$

If, for brevity, we denote the Hessian matrix of $W$ with respect
to the $v^a_i$ by $\calH$, then $\zeta$ is the vector field along
${\cal C}$ given by
$$
  \zeta = (\calH^{-1})^{ab}_{ij} {\mathcal J} (v^{-1})^j_b
  \frac{\partial}{\partial v^a_i}.
$$
Let us consider the function $f:=\zeta(\varphi)$, or explicitly
$$
  f = (\calH^{-1})^{ab}_{ij} {\mathcal J}^2 (v^{-1})^i_a (v^{-1})^j_b\,.
$$
For each $\gamma \in {\cal C}$, $f(\gamma) \neq 0$ from which it
follows that the compatibility condition (\ref{compatibility})
holds. The nonholonomic projector ${\cal P}$ is then found to be
$$
  \calP = I - f^{-1} d \varphi \otimes \zeta\,.
$$

\section{Cauchy formalism for nonholonomic field theory}\label{par:cauchy}

We will now describe the transition from the multisymplectic
(covariant) treatment of nonholonomic field theory, discussed in
the previous sections, to the formulation of the problem on the
space of Cauchy data. The Cauchy formalism for field theories is
an infinite-dimensional analogue of classical dynamics of systems
with a finite number of degrees of freedom. Instead of looking for
sections of a bundle $Y$ over an $(n+1)$-dimensional space-time
manifold $X$ (as in the covariant approach), one starts by
introducing a space $\tX$ of embeddings of a fixed `Cauchy
surface' into $X$. This space replaces the absolute time from
Newtonian mechanics and, under suitable conditions, the system can
then be described in terms of a particular vector field on an
infinite-dimensional manifold $\tZ$, called the space of Cauchy
data, which is a bundle over $\tX$.

In our discussion of the Cauchy formalism for nonholonomic field
theory, attention will be focussed on the case where the base
manifold $X$ admits a global splitting in `space' and `time'. It
will be shown that, under appropriate assumptions, this Cauchy
formalism reveals a close resemblance to the cosymplectic
formulation of time-dependent nonholonomic mechanics (see e.g.\
\cite{bracket00}). This is in agreement with the results described
in \cite{santamaria04} for unconstrained Lagrangian field theory.
Our aim is mainly to present the general idea, without entering
into all technical details related to the geometry and analysis on
infinite dimensional manifolds.

Finally, it should be noted that only for \textit{hyperbolic}
partial differential equations it makes sense to consider initial
value problems. In the remainder of this section we will therefore
tacitly assume that the field equations we are dealing with, are
hyperbolic in some suitable sense. We refer to
\cite{christodoulou} for a detailed analysis of this matter.

\subsection{The space of Cauchy data}
We first recall some basic aspects of the Cauchy formalism for
Lagrangian field theories. We thereby closely follow the
treatments presented in \cite{bsf88,symm04,santamaria04}, to which
we also refer for more details and further references on the
subject.

\subsubsection{Generalities}
As before, we start from a fibre bundle $\pi: Y \rightarrow X$
whose base space $X$ is an $(n + 1)$-dimensional orientable
manifold. Let $M$ be an $n$-dimensional compact oriented manifold
with volume form $\eta_M$. The pair $(M,\eta_M)$ is called a
Cauchy surface. A \emph{space of (parametrized) Cauchy surfaces}
$\tX$ is then defined as a smooth manifold of embeddings $\tau: M
\hookrightarrow X$.
\begin{remark}
Usually, $X$ and $M$ are taken to be manifolds with boundary and
the embeddings $\tau$ belonging to $\tX$  are then assumed to map
the interior, resp.\ boundary, of $M$ into the interior, resp.\
boundary, of $X$. However, for the purpose of the present paper we
will leave all considerations related to boundary aspects aside.
\end{remark}
In the sequel we will always assume, without loss of generality,
that $M$ has volume one, i.e.
\begin{equation}\label{volume}
\int_M\eta_M = 1\,.
\end{equation}
Points of $M$ will usually be denoted by $u$.

Given a space of Cauchy surfaces $\tX$, the \emph{space of Cauchy
data} $\tZ$ is defined as a (infinite dimensional) manifold of
embeddings from $M$ into $J^1\pi$, having the property that for
each embedding $\kappa: M \hookrightarrow J^1\pi$, there exists a
section $\phi$ of $\pi$ and an element $\tau$ of $\tX$ such that
$\kappa = j^1\phi \circ \tau$. Finally, we define the \emph{space
of Dirichlet data} $\tY$ as consisting of those embeddings
$\delta: M \hookrightarrow Y$ having the property that there
exists an element $\kappa$ of $\tZ$ such that $\delta = \pi_{1, 0}
\circ \kappa$.

It is obvious from the previous definitions that the respective
projections $\pi_{1, 0}: J^1\pi \rightarrow Y$ and $\pi: Y
\rightarrow X$ induce the following natural projections:
$$
   \tZ \stackrel{\tpi_{1,0}}{\longrightarrow} \tY
      \stackrel{\tpi}{\longrightarrow} \tX.
$$
We further put $\tpi_1 = \tpi \circ \tpi_{1,0}$: the projection of
$\tZ$ onto $\tX$.

The spaces $\tX$, $\tY$, and $\tZ$ can be equipped with the
Whitney topology and can be made into (infinite-dimensional)
manifolds (see \cite{bsf88}).  Because of the compactness of $M$,
it does not matter whether one chooses the strong or the weak
Whitney topology. However, as mentioned at the beginning of this
section, we will not unduly concern ourselves with technical
issues related to the smooth nature of these manifolds and of the
mappings and other objects defined on them. For a detailed
discussion of the topology and differentiable structure on a space
of differentiable mappings between manifolds, we refer to
\cite{michor}.

The tangent space of $\tX$ has a convenient geometrical
interpretation (see \cite{bsf88, gimmsyII}): let $\tau \in \tX$,
then $T_\tau \tX$ can be identified with the space of sections of
the pull-back bundle $\tau^\ast TX$. Equivalently, a vector
$V_\tau \in T_\tau \tX$ can be identified with a vector field
along $\tau$, i.e. $V_\tau: M \longrightarrow TX,\,u\longmapsto
V_\tau(u) \in T_{\tau(u)} X$. Since $\tau$ is a bijection (onto
its image), one can still identify $V_\tau$ with a vector field on
$X$, defined along $\tau(M)$. Similar interpretations exist for
elements of $T\tY$ and $T\tZ$ and will be freely used in the
sequel. For instance,  given $\kappa \in \tZ$, we will use the
notation $W_{\kappa}$ to indicate both an element of the tangent
space $T_{\kappa}\tZ$ and the corresponding vector field on
$J^1\pi$ along $\kappa (M)$.

\subsubsection{Existence of a global splitting of $X$}
For the further discussion we assume that there exists a global
splitting of the base manifold $X$, induced by a diffeomorphism
$\Psi: \bfR \times M \rightarrow X$. In physics, $\bfR$ is usually
associated to time and $M$ to (physical) space. An embedding
$\tau: M \hookrightarrow X$ is then called \emph{admissible} if
there exists a (necessarily unique) $t \in \bfR$ such that $\tau
(u) = \Psi(t,u)$. We henceforth restrict $\tX$ to be a manifold of
admissible embeddings of $M$ in $X$. This has the effect of
reducing $\tX$ to a $1$-dimensional space diffeomorphic to $\bfR$,
with coordinate function denoted by $t$. The spaces of Cauchy data
$\tZ$ and of Dirichlet data $\tY$, are then restricted
accordingly. There is a canonically defined vector field $\Xi$ on
$\tX$, given by
$$
  \Xi(\tau)(u) = \frac{d}{ds}\Psi(s, u)\Big|_{s = t} \quad
  \text{for all $u \in M$},
$$
where $t$ is such that $\tau (\cdot) = \Psi(t, \cdot)$. In
particular we have that $\langle \Xi,dt\rangle = 1$. The global
time slicing of $X$ allows us to consider a volume form $\eta$ on
$X$ which, with some abuse of notation, can be written as
\[ \eta
= dt \wedge \eta_M\,.
\]
(see e.g.\ \cite{santamaria04}). In the sequel we will always
assume that $X$ is oriented in terms of this volume form.

An important property now is that, under the above assumptions
(i.e.\ existence of a global splitting of $X$ and $\tX$ consisting
of admissible embeddings only) one can prove that the space of
Cauchy data $\tZ$ is diffeomorphic to the jet bundle $J^1\tpi$
(see \cite[par. 5.2.]{santamaria04}). To make this diffeomorphism
more explicit we note that there is a one-to-one correspondence
between sections $\phi$ of $\pi_1$ and sections $\varphi$ of
$\tpi_1$:
$$
  \phi(x) = \varphi(\tau)(u), \quad \text{where $x = \Psi(t, u)$ and
    $\tau (\cdot)= \Psi(t, \cdot)$}.
$$
We will frequently switch back and forth between both
interpretations without warning, but we will stick to the notation
``$\phi$'' for a section of $\pi_1$ and ``$\varphi$'' for the
corresponding section of $\tpi_1$.

\subsection{The unconstrained Lagrangian formalism}
Starting from a Lagrangian density $L\eta$ on $J^1\pi$, with
regular Lagrangian $L$,  the multisymplectic $(n+2)$-form
$\Omega_L$ induces a $2$-form $\tOmega_L$ on the space $\tZ$ of
Cauchy data as follows. Let $\kappa \in \tZ$, and
$W_\kappa,W'_\kappa \in T_\kappa\tZ$, then put
$$
  \tOmega_L(\kappa)(W_\kappa,W'_\kappa) :=
  \int_M \kappa^\ast( i_{W_\kappa}\, i_{W'_\kappa} \Omega_L)\,,
$$
where on the right-hand side, $W_\kappa$ and $W'_\kappa$ are
interpreted as vector fields on $J^1\pi$, defined along
$\kappa(M)$ (see the end of subsection 6.1.1). Likewise, the
$(n+1)$-form $\eta$ (pull-back of the volume form on $X$) induces
a one-form $\teta$ on $\tZ$ according to the prescription
\[ \teta (\kappa)(W_\kappa) := \int_M
\kappa^*(i_{W_\kappa} \eta)
\]
for all $\kappa \in \tZ,\, W_\kappa \in T_\kappa\tZ$. One can
prove that both $\tOmega_L$ and $\teta$ are closed forms and, in
particular, it turns out that $\tOmega_L = - d\tilde{\Theta}_L$,
where $\tilde{\Theta}_L$ is the one-form on $\tZ$ induced by
$\Theta_L$ (cf. \cite{santamaria04} for more details).

As for the jet bundle $J^1\pi$, one can show that the space of
Cauchy data $\tZ$ can be equipped with a `vertical endomorphism'
$\tS_{\tilde{\eta}}$ (see \cite[section 5.2.3]{santamaria04}). In
the case under consideration, with $\tX$ being 1-dimensional,
$\tS_{\tilde{\eta}}$ is a vector valued one-form that can be
defined as follows. Take any $\kappa \in \tZ$, with $\kappa =
j^1\phi \circ \tau$ for some $\tau \in \tX$ and section $\phi$ of
$\pi$. In view of the identification between $\tZ$ and $J^1\tpi$
(see subsection 6.1.2), we can still represent $\kappa$ by
$j^1_\tau\varphi$. For arbitrary $W_\kappa \in T_\kappa \tZ$, we
then put
\begin{equation}\label{endo}
  \tS_{\tilde{\eta}}(W_\kappa) = \left( T_{j^1_\tau \varphi} \tpi_{1,0}(W_\kappa) - T_\tau
  \varphi \circ T_{j^1_\tau \varphi} \tpi_1(W_\kappa) \right)^v\,,
\end{equation}
where the superscript `$v$' denotes the natural vertical lift
operation from $T\tY$ into $V\tpi_{1,0}$. With the terminology
used for vector fields on a first-order jet bundle, we will say
that a vector field $\Gamma$ on $\tZ$ is \emph{a second-order
vector field} (shortly, a SODE) if
\begin{equation}\label{sode}
\tS_{\tilde{\eta}}(\Gamma) = 0\quad \mbox{and}\quad
i_{\Gamma}\teta = 1\,.
\end{equation}

Consider a connection (or jet field) $\Upsilon$ on $\pi_1: Y
\longrightarrow X$, with horizontal projector $\proj$. One can
then construct a vector field $\Gamma$ on $\tZ$ as follows. For
$\kappa \in \tZ$, with $\kappa = j^1\phi \circ \tau$, define the
vector $\Gamma (\kappa) \in T_\kappa\tZ$ by
\begin{equation}\label{gamma}
  \Gamma(\kappa)(u) =
    \proj\left( T j^1\phi( \Xi(\tau)(u) )\right),
\end{equation}
i.e.\ $\Gamma (\kappa)(u) \in T_{\kappa (u)}J^1\pi$ is the
horizontal lift of $\Xi (\tau)(u) \in T_{\tau (u)}X$ under the
given connection $\Upsilon$. We then have the following
interesting property.

\begin{prop} \label{prop:sode}
If $\Upsilon$ is a semi-holonomic connection on $\pi_1$, then the
vector field $\Gamma$ on $\tZ$, defined by (\ref{gamma}) is a
second-order vector field.
\end{prop}
\begin{mproof}
For the contraction of $\Gamma$ with $\teta$ we find that
$$
  \left( i_\Gamma \teta \right)(\kappa) = \int_M
  \kappa^\ast(i_{\Gamma(\kappa)} \eta) = \int_M
  \tau^\ast(i_{\Xi(\tau)} \eta) = 1\,,
$$
where the last equality follows from the normalization assumption
(\ref{volume}) and for the second equality we have used the fact
that (with previous conventions) $i_{\Gamma(\kappa)} \eta =
\pi^*_1\left(i_{\Xi(\tau)} \eta\right)$ and $\pi_1 \circ \kappa =
\tau$. Herewith, we have already shown that $\Gamma$ verifies the
second condition of (\ref{sode}).

Next, we investigate the first condition of (\ref{sode}).  Since
the given connection $\Upsilon$ is semi-holononomic, it is easily
checked in coordinates that $\proj$ satisfies
\begin{equation}\label{sh} T_\gamma
  \pi_{1,0}( \proj(v_\gamma)) = T_\gamma(\phi \circ \pi_1)(v_\gamma),
\end{equation}
where $\gamma = j^1_x\phi$ and $v_\gamma \in T_\gamma J^1\pi$. We
now compute $\tS_{\teta}(\Gamma(\kappa))$. With $W_\kappa = \Gamma
(\kappa)$, the first term on the right-hand side of (\ref{endo})
becomes
\begin{eqnarray*}
  T_\kappa \tpi_{1,0}( \Gamma(\kappa) ) (u) & = &
    T_{\kappa(u)} \pi_{1,0}( \Gamma(\kappa)(u) ) \\
    & = & T_{\kappa(u)} \pi_{1,0}(\proj(K_{\kappa(u)})),
\end{eqnarray*}
where $\kappa = j^1_\tau \varphi$ and where, for notational
convenience, we have abbreviated $T j^1\phi(\Xi(\tau)(u))$ by
$K_{\kappa(u)}$. Using property (\ref{sh}), we further obtain
\begin{eqnarray*}
  T_\kappa \tpi_{1,0}( \Gamma(\kappa) ) (u) & = &
    T_{\kappa(u)}(\phi \circ \pi_1)( K_{\kappa(u)} ) \\
    & = & T_{\kappa(u)} \phi (\Xi(\tau)(u)),
\end{eqnarray*}
so that
$$
  T_\kappa \tpi_{1,0}( \Gamma(\kappa) ) = T_\tau \varphi(\Xi(\tau)),
$$
from which it follows that $\tS_{\teta}(\Gamma(\kappa)) = 0$,
which completes the proof that $\Gamma$ defines a second-order
ODE.\QED\end{mproof}
We then arrive at the following important
result in the Cauchy formalism for (unconstrained) Lagrangian
field theory.

\begin{theorem} \label{prop:inftyDDW}
If $\proj$ satisfies the De Donder-Weyl equation (\ref{DDW}), then
the vector field $\Gamma$ on $\tZ$, defined by (\ref{gamma}),
satisfies the equations
\[
i_\Gamma \tOmega_L = 0 \quad \mbox{and} \quad i_\Gamma\teta = 1\,.
\]
\end{theorem}
\begin{mproof} See \cite[chapter 5]{santamaria04}.\QED
\end{mproof}

\subsection{Nonholonomic constraints}
We now return to the nonholonomic setting described in
sections~\ref{par:nh} and \ref{par:nhproj} and, in particular, we
assume that the compatibility condition (\ref{compatibility})
holds. In order to adapt the Cauchy formalism, discussed in the
previous subsection, to the nonholonomic case, we first define a
subset $\tcalC$ of $\tZ$ as follows:
$$
  \tcalC := \left\{ \kappa \in \tZ \, |\, \im \kappa \subset \calC
  \right\}.
$$
This set can be equipped with a smooth manifold structure such
that $\tcalC$ becomes a (infinite-dimensional) submanifold of
$\tZ$. The tangent space to $\tcalC$ at a point $\kappa$ is given
by $T_\kappa\tcalC = \{W_\kappa \in T_\kappa\tZ\,|\, W_\kappa (u)
\in T_{\kappa(u)}\calC\}$.

For each $\kappa \in \tcalC$, let
$$
  \tD_\kappa := \left\{ W_\kappa \in T_\kappa \tZ \,|\; \im W_\kappa
  \subset D \right\}\,,
$$
where $D$ is the constraint distribution along $\cal C$. Putting
\[
\tD = \bigcup_{\kappa \in\, \tcalC} D_\kappa
\]
one may verify that $\tD$ determines a smooth distribution on
$\tZ$ along $\tcalC$.

Next, for $\kappa \in \tcalC$ and for each section $\alpha$ of the
bundle $F$ of constraint forms along ${\cal C}$, we define an
element $\talpha_\kappa$ of $T_\kappa^*\tZ$ by
\[
  \talpha_\kappa( W_\kappa ) = \int_M \kappa^\ast( i_{W_\kappa} \alpha ),
     \quad \text{for all $W_\kappa \in T_\kappa \tZ$}.
\]
The set of all such covectors $\talpha_\kappa$ determines a
subspace $\tF_\kappa$ of  $T_\kappa^*\tZ$ and
\[
\tF = \bigcup_{\kappa \in\, \tcalC} \tF_\kappa
\]
is a codistribution on $\tZ$ along $\tcalC$.

Since we assume that the given constrained problem satisfies the
compatibility condition, we can use the nonholonomic projector
$\cal P$ and the complementary projector ${\cal Q} = I - {\cal P}$
(cf. Section \ref{par:nhproj}) to define two operators
${\tcalP},\,{\tcalQ}: T_{\tcalC}\tZ \longrightarrow T_{\tcalC}\tZ$
as follows. For each $\kappa \in {\tcalC}$ and $W_\kappa \in
T_\kappa \tZ$, put
\[
    \tcalP_\kappa(W_\kappa) = \calP \circ
    W_\kappa\; (\in T_\kappa\tZ), \qquad \tcalQ_\kappa (W_\kappa)=\calQ \circ W_\kappa\; (\in T_\kappa\tZ)\,.
\]
Using the properties of ${\cal P}$ and ${\cal Q}$, it is not hard
to check that, for each $\kappa \in \tcalC$, $\tcalP_\kappa$ and
$\tcalQ_\kappa$ define complementary projectors in $T_\kappa \tZ$,
i.e.
$$
  (\tcalP_\kappa)^2 = \tcalP_\kappa,\; (\tcalQ_\kappa)^2 = \tcalQ_\kappa
   \quad \text{and}\quad \tcalP_\kappa + \tcalQ_\kappa =
   I_\kappa\,,
$$
with $I_\kappa$ the identity on $T_\kappa\tZ$. This implies that
$T_\kappa\tZ = \im \tcalP_\kappa \oplus \im \tcalQ_\kappa$. Again
relying on the definitions of $\tcalC, \tD, \tcalP$ and $\tcalQ$,
and on the properties of the nonholonomic projector $\calP$, one
can prove that
$$
    \im \tcalP_\kappa = T_\kappa \tcalC \quad \text{and} \quad
      \im \tcalQ_\kappa = \tD_\kappa.
$$
Summarizing, we may conclude that under the given conditions we
have the following decomposition of $T\tZ$ along $\tcalC$:
\[
 T_{\tcalC}\tZ = T\tcalC \oplus \tD\,.
\]

Let $\proj$ be the horizontal projector of a connection $\Upsilon$
on $\pi_1$ and let $\Gamma$ denote the vector field on $\tZ$
defined by (\ref{gamma}). The composition $\tcalP \circ
\Gamma_{|\tcalC}$ then determines a vector field on $\tcalC$,
shortly denoted by $\tcalP (\Gamma)$, and it is not difficult to
see that it is precisely the vector field associated to the
induced connection on $(\pi_1)_{|\calC}$ with horizontal projector
${\proj}_{\calP} = \calP \circ \proj$ (see Section
\ref{par:nhproj}). We now have the following interesting result.
\begin{lemma}
There exists a section $\talpha$ of $\tF$, such that
\begin{equation}\label{inftyproj}
  i_{\tcalP(\Gamma)} \tOmega_L = i_\Gamma \tOmega_L + \talpha.  \end{equation}
\end{lemma}
\begin{mproof}
For $\kappa \in \tcalC$ and $W_\kappa \in T_\kappa \tZ$, one can
deduce from the definition of $\tOmega_L$ that
$$
  (i_{\tcalP(\Gamma)} \tOmega_L)(\kappa)(W_\kappa) = \int_M \kappa^\ast(
    i_{\tcalP(\Gamma)(\kappa)} i_{W_\kappa} \Omega_L).
$$
For the integrand on the right-hand side we have that, with $u \in
M$,
$$
  i_{\tcalP(\Gamma)(\kappa)(u)} i_{W_\kappa(u)} \Omega_L =
    i_{\Gamma(\kappa)(u)} i_{W_\kappa(u)} \Omega_L
    - i_{\tcalQ(\Gamma)(\kappa)(u)} i_{W_\kappa(u)}\Omega_L,
$$
where $\tcalQ(\Gamma)$ is the vector field associated to $\calQ
\circ \proj$ (note that $\tcalQ(\Gamma)$ is defined along $\tcalC$).
Since $\tcalQ(\Gamma)(\kappa)(u)$ is an element of the constraint
distribution $D$, the contraction with $\Omega_L$ yields a form
$\alpha_{\kappa (u)} \in F_{\kappa (u)}$. Integration over $M$
then gives (\ref{inftyproj}).\QED
\end{mproof}
We have now collected all ingredients needed to formulate the main
result of this section. Consider a constrained Lagrangian field
theory, with regular Lagrangian $L$, with constraints verifying
the appropriate conditions and such that the base manifold $X$
admits a global space-time splitting.
\begin{theorem} \label{prop:indconstr}
Let $\proj$ be a solution of the unconstrained De Donder-Weyl
equation (\ref{DDW}) and let $\Gamma$ be the corresponding
second-order vector field on $\tZ$. Then, the vector field
$\tcalP(\Gamma)$ on $\tcalC$ satisfies the following relations:
  \begin{equation}\label{constrinftyDDW}
    i_{\tcalP(\Gamma)} \tOmega_L \in \tF \text{ and }
      \tcalP(\Gamma) \in T \tcalC.
  \end{equation}
\end{theorem}
\begin{mproof}
If $\proj$ satisfies the De Donder-Weyl equation, then the
associated vector field $\Gamma$ is contained in the kernel of
$\tOmega_L$ (see proposition \ref{prop:inftyDDW}).  Expression
(\ref{inftyproj}) then proves the first part of
(\ref{constrinftyDDW}).  The second part follows from the
definition of $\tcalC$.\QED
\end{mproof}
In addition, we note that $\tcalP(\Gamma)$ is still a vector field
of second-order type, due to propositions \ref{lemma:semihol} and
\ref{prop:sode}.

To conclude, we have shown that under the appropriate assumptions,
the Cauchy formalism for nonholonomic field theory leads to a
vector field of `second-order type' on the (infinite dimensional)
subspace $\tcalC$ of the space of Cauchy data $\tZ$, which can be
written as a projection of the second-order vector field on $\tZ$
associated to the free (unconstrained) Lagrangian system.

\section{Some final comments}
In this paper we have studied various aspects of nonholonomic
Lagrangian field theory. Among others, we have shown that both in
the multisymplectic approach and in the Cauchy formalism, the
equations for the constrained system can be obtained by a
projection of the equations for the original unconstrained
Lagrangian system.

While finalizing this paper we have come across a recent work of
Olga Krupkova (\cite{krupkova}) in which nonholonomic Lagrangian
field theory is discussed within the framework of a general study
of partial differential equations with differential constraints.
This paper --- which differs both in purpose and methodology from
ours --- presents, among others, an interesting analysis of the
various types of constraints that one may encounter when dealing
with constrained exterior differential systems on fibred
manifolds.

The subject of nonholonomic field theory is still in full
development. As far as the present study is concerned, there still
remains some work to be done concerning the Cauchy formalism,
mainly regarding the technicalities related to the
infinite-dimensional manifold structure of the spaces $\tZ$ and
$\tcalC$. Another interesting matter that will be treated in
future work, concerns the comparison between the constrained
variational approach and the nonholonomic approach, i.e.\ the
field theoretic analogue of the comparison between vakonomic and
nonholonomic mechanics (see e.g.\ \cite{cortes2}).

Finally, an important challenge for future work will be the
identification of some physically relevant examples to which
nonholonomic field theory can be applied and for which, unlike the
example of incompressible hydrodynamics treated in Section 5, the
nonholonomic and the constrained variational approach are not
``equivalent". It is to be expected that interesting examples
should come, for instance, from problems in elasticity (such as
the rolling without slipping of a deformable body over a surface).

{\bf Acknowledgements}\\
J.V and F.C wish to thank the Research Foundation-Flanders (FWO)
and the ``Bijzonder Onderzoeksfonds" (BOF) of Ghent University, for
financial support. They also thank the CSIC (Madrid) for its kind
hospitality on the occasion of several research visits. M.dL and
D.MdD wish to thank  to MICYT (Spain) Grant MTM2004-7832 for
financial support. We would like to thank Marcelo Epstein  for
valuable discussions.

\end{document}